\RequirePackage{fix-cm}
\documentclass[twocolumn]{svjour3}          
\smartqed  

\usepackage{graphicx}
\usepackage{multirow}
\usepackage{color}
\usepackage{amsmath}
\usepackage{graphicx}
\usepackage{float}
\usepackage{subfig}
\usepackage{amssymb}

\usepackage{algorithm}
\usepackage{algpseudocode}

\usepackage{url}

\journalname{}
\begin{document}
\title{Transaction Fee Estimation in the Bitcoin System
}


\author{Limeng Zhang         \and
        Rui Zhou  \and
        Qing Liu \and
        Chengfei Liu \and
        M. Ali Babar
}


\institute{Limeng Zhang \at
              Swinburne University of Technology, Melbourne, Australia \\
              \email{limengzhang@swin.edu.au}           
           \and
           Rui Zhou \at
              Swinburne University of Technology, Melbourne, Australia \\
              \email{rzhou@swin.edu.au}
           \and
           Qing Liu \at
           Data61, CSIRO, Hobart, Australia\\
           \email{q.liu@data61.csiro.au}
           \and
           Chengfei Liu \at
           Swinburne University of Technology, Melbourne, Australia\\
           \email{cliu@swin.edu.au}
           \and
                      M.Ali Babar \at
           The University of Adelaide, Adelaide, Australia\\
           \email{ali.babar@adelaide.edu.au}
}

\date{Received: date / Accepted: date}
\maketitle
\begin{abstract}
In the Bitcoin system, transaction fees serve as an incentive for blockchain confirmations. In general, a transaction with a higher fee is likely to be included in the next block mined, whereas a transaction with a smaller fee or no fee may be delayed or never processed at all.   However, the transaction fee needs to be specified when submitting a transaction and almost cannot be altered thereafter. Hence it is indispensable to help a client set a reasonable fee, as a higher fee incurs over-spending and a lower fee could delay the confirmation.  In this work, we focus on estimating the transaction fee for a new transaction to help with its confirmation within a given expected time.  We identify two major drawbacks in the existing works. First, the current industry products are built on explicit analytical models, ignoring the complex interactions of different factors which could be better captured by machine learning based methods; Second, all of the existing works utilize limited knowledge for the estimation which hinders the potential of further improving the estimation quality. As a result, we propose a framework FENN, which aims to integrate the knowledge from a wide range of sources, including the transaction itself, unconfirmed transactions in the mempool and the blockchain confirmation environment, into a neural network  model in order to estimate a proper transaction fee. Finally, we conduct experiments on real blockchain datasets to demonstrate the effectiveness and efficiency of our proposed framework over the state-of-the-art works evaluated by MAPE and RMSE. Each variation model in our framework can finish  training  within one block interval, which shows the potential of our framework  to process the realtime transaction updates in the Bitcoin blockchain.  


\keywords{Transaction Fee Estimation \and Bitcoin \and Blockchain \and Transaction Confirmation}
\end{abstract}

\section{Introduction}
\label{intro}

Nowadays cryptocurrencies  have become a buzzword in both industry and academia \cite{mukhopadhyay2016brief,wood2014ethereum,schwartz2014ripple}, and  the market cap of Bitcoin, one of the most popular cryptocurrencies,  reached a record high of over 1,200 billion USD in November 2021\footnote{https://www.slickcharts.com/currency}. Blockchain provides the techniques underpinning for managing cryptocurrencies  through a distributed ledger. Compared to the traditional currency systems \cite{jack2011mobile,chaum1997david}, the blockchain distributed ledger has three novel characteristics: decentralization, transparency and immutability.  Currently, the majority of blockchain research focuses on the technologies that underpinning the deployment of blockchain systems, such as  consensus protocol \cite{eyal2016bitcoin,gilad2017algorand,luu2016secure,zheng2017overview,vukolic2015quest}, query evaluation \cite{zhang2019gem,zhu2019sebdb,xu2019towards,xu2019vchain}, scalability enhancement \cite{xu2018cub,sharma2019blurring,dang2019towards}, system engineering \cite{dinh2017blockbench,dinh2018untangling,wang2018forkbase,ruan2019fine}, etc. However, one component is also indispensable  to have a cryptocurrency system to be widely acceptable or even more prominent. That is the usability of the system-affiliated software, eg., client-side software.

In this paper, we focus on one important usability-related issue, transaction fee estimation in the  Bitcoin blockchain system. The transaction fee serves as an incentive to complete the transfer of Bitcoin assets (recorded in a transaction) on the blockchain \cite{easley2019mining,kasahara2016effect,li2018transaction,eyal2014majority}. Generally, a transaction with a higher fee  is likely to be processed faster, whereas a transaction with a lower fee or no fee may be delayed for long  or may never be processed. However, a user is required to specify a transaction fee when the user submits the transaction. 
If the specified fee is too high, the user will suffer from over-spending; if the specified fee is too low, the transaction may not be confirmed in time. Unfortunately, a user usually has no idea about how to properly preset a fee. We strongly believe the problem needs an in-depth study. A helpful fee estimation software could attract a large number of users and win a substantial portion of the market.

Most of the existing fee estimation tools were developed as built-in components of Bitcoin software by industry developers, such as BtcFlow\footnote{https://bitcoiner.live/\label{BtcFlow}}, Bitcoin Core\footnote{https://bitcoin.org/en/bitcoin-core/\label{BCore}}, statoshi\footnote{https://statoshi.info/dashboard/db/fee-estimates/\label{statoshi}}, buybitcoinworldwide\footnote{https://www.buybitcoinworldwide.com/fee-calculator/\label{buybitcoinworldwide}}, bitcoinfees\footnote{https://bitcoinfees.earn.com/\label{bitcoinfees}}, etc.
All of these strategies are based on the assumption that transactions with greater transaction fees are superior during the confirmation process.
BtcFlow returns the minimal transaction fee for transactions in which the arrival rate is equal to or less than the confirmation rate. Bitcoin Core\textsuperscript{\ref{BCore}}  delivers a weighted transaction fee based on the transaction fees of blockchain transactions with the same confirmation time. This result also applies to  Statoshi\textsuperscript{\ref{statoshi}} and buybitcoinworldwide\textsuperscript{\ref{buybitcoinworldwide}}. Bitcoinfees\textsuperscript{\ref{bitcoinfees}} analyses the confirmation possibility of different transaction fees through an estimation API, which is not open sourced. As a result, it only works on real time online data and could not be evaluated and compared with other methods on collected datasets.

Although the existing industry tools have provided preliminary solutions to the fee estimation problem, there are opportunities lying in the following three aspects. (1) \textbf{Insufficient disclose of the industry methods to academic community}: Industry methods are not systematically documented and published academically (even for those open source tools), so researchers are not much aware of them, not to mention thinking of improving the existing methods. (2) \textbf{Inferior estimation accuracy}: Most of the existing methods are built  on explicit analytical models, which means they consider transaction confirmation is driven by fixed mechanisms, eg., higher fee strictly affirms earlier confirmation in BtcFlow\textsuperscript{\ref{BtcFlow}} and Bitcoin Core\textsuperscript{\ref{BCore}}, blocks are generated following a specific Poisson distribution in BtcFlow\textsuperscript{\ref{BtcFlow}}, historical confirmation time determines future confirmation time in Bitcoin Core\textsuperscript{\ref{BCore}}. The existing methods ignore the complex interactions of different factors which could be better captured by machine learning based methods. (3) \textbf{Limited knowledge is utilized}: Partly caused by the intrinsic characteristics of the analytical models, it is difficult to integrate a diverse range of knowledge sources to do the estimation, eg., BtcFlow\textsuperscript{\ref{BtcFlow}} mainly makes use of the information of newly submitted transactions in mempool, Bitcoin Core mainly considers historical transaction fees recorded in the blockchain. Surprisingly, the details of a given transaction itself are not sufficiently considered, eg., transaction inputs and outputs are two essential factors contributing to the complexity of transaction verification \cite{antonopoulos2014mastering}. Other factors include block capacity (as it controls the upper bound of  the transaction volume in each block), the blockchain network (such as computational power) evolving mining efficiency (which in return affects the speed of transaction confirmation), etc.

Other than the industry products, AI-Shehabi \cite{al2018bitcoin} proposed a neural network framework to model the interactions of transaction fee and transaction confirmation time. By assuming that the unconfirmed transactions in the mempool comprise the future block sequence in the blockchain, it organises these transactions, according to their transaction fees, into a virtual sequence of blocks. It then uses neural networks to extract patterns among  the transaction fee, the virtual confirmation time (virtual block position), and its actual confirmation time, which are then used for fee prediction.

To further investigate the transaction fee estimation problem, we construct a new framework integrating more details in transactions and blocks. In our framework, we take into account, the features of the transaction itself, unconfirmed transactions in the mempool and the blockchain confirmation environment, such as the mining rate, block capacity, etc.  We tackle this problem by leveraging the power of neural networks. Specifically, we make the following contributions
\begin{itemize}
  \item We organize and systematically document the principle and the detailed estimation procedure of existing transaction fee estimation works from industry tools. Meanwhile, we analyse the limitations of these works. Finally, we implemented them to support offline evaluation. 
\item  We propose a  neural network framework to capture the complex interactions of different factors related to transaction confirmation. Meanwhile, compared to information used in the existing works, we integrate  the  knowledge from  a  more wide  range  of  sources,  including  the transaction itself, unconfirmed transactions in the mempool and the blockchain confirmation environment, in order to estimate a proper transaction fee.

\item  We conduct extensive experiments on the real-world datasets to study the performance of fee estimation algorithms to demonstrate the efficiency and effectiveness of our proposed framework.

\end{itemize}

The rest of this paper is structured as follows: We introduce blockchain background information in Section \ref{secBackground} and define the target research problem in Section \ref{sectProblemDefinition}. Related works on transaction fee estimation are studied in Section \ref{secBtcFlow}-\ref{secMSLP}, where we assess the benefits and limitations of the existing fee estimation methods. In  Section \ref{secFENN}, we present the proposed transaction fee estimation framework. In the following Section \ref{secExperiments}, we conduct experiments to evaluate the performance of different fee estimation methods as well as our proposed solution. Finally, we conclude our work in Section \ref{secConclusion}.

\section{Preliminaries}
\label{secBackground}
Proposed in 2008, the Bitcoin blockchain system as a decentralized digital currency payment system operates on a worldwide basis \cite{nakamoto2008bitcoin}. Even though different cryptocurrencies, such as Ethereum (ETH) \cite{wood2014ethereum}, Dash\footnote{https://www.dash.org/}, Ripple (XRP) \cite{schwartz2014ripple}, Litecoin\footnote{https://litecoin.org/} (LTC),  have been designed, a recent study shows that Bitcoin remains the dominant cryptocurrency in terms of market capitalisation and it is  the most widely supported cryptocurrency among participating exchanges, wallets and payment companies \cite{hileman2017global,mukhopadhyay2016brief}. Blockchain revolutionizes the way we interact, automate payments, trace and track transactions. Fundamentally, it uses an immutable linked chain of blocks to record and track transactions. \cite{nakamoto2008bitcoin,salah2019blockchain,yaga2019blockchain}.

  Transactions  in the Bitcoin system record the digital asset transfer between  clients. In a transaction, the output describes the amount of digital assets to be transferred to the new owner(s), while the input identifies the digital assets to be consumed \cite{nakamoto2008bitcoin}.  One transaction can have several inputs and outputs. \textit{Transaction fee} is set with the difference between the total of input and output assets, which will be collected by miners once the transaction is confirmed. Consequently, miners often choose transactions with larger transaction fees to maximise their mining profits.   As a result, the transaction fee has to be increased to help with transaction confirmation by increasing the processing priority.  \textit{Transaction feerate} measures the fee density of a transaction or the transaction fee per size unit. In fact, each time only a limited number of transactions can be confirmed, hence miners would prioritise transactions with higher feeratess in order to increase mining profits.

 Transactions  can be submitted to the blockchain network via the Bitcoin wallets (desktop wallet, mobile wallet, web wallet, paper wallet, etc.) \cite{nakamoto2008bitcoin}. The submitted pending transactions  are then broadcast across  various nodes.  If they meet the transactions' validity criteria \cite{antonopoulos2014mastering}, Bitcoin nodes will add them to their mempools (memory pools), where transactions wait until they can be included (mined) into a block. Miners are the nodes responsible for the verification and block construction. Specifically, miners first select unconfirmed transactions from the mempool to construct their own candidate blocks, and then compete to determine a computational solution (Proof of Work (Pow)) to attach their candidature block to the blockchain (referred to as `mining a block'). Among all the miners, only the first miner who solves this computational issue will be rewarded with a fixed Bitcoin reward from the blockchain system plus transaction fees from the chosen transactions. Once a new block is added to the blockchain, these transactions are confirmed and will then be removed from the miner's mempool.

A transaction is complete when it is recorded in a blockchain block. The block  stores information about the selected transactions, as well as information about the mining complexity, block size, etc. In the meantime, the Bitcoin blockchain system keeps track of the generation time of each block.

\section{Problem definition}
\label{sectProblemDefinition}
The paper targets on estimating the transaction fee for a transaction with a given expected confirmation time (time  between entering mempool and confirmed in the blockchain). As for the starting timestamp of confirmation time, a more realistic and meaningful choice could be the submission time for a transaction. We make our decision on using the entering mempool time mainly for two reasons. The first one is that the submission timestamp is unavailable. The second one is that the entering time is more reasonable. In the blockchain system, there can be different kinds of time delays between transaction submission and transaction entering the mempool, such as the propagation routine among network nodes, network traffic, etc. However, once the transactions enter the mempool, it means that they will start to compete to be included in the next block.

Specifically, the estimation process may rely on  three kinds of data, namely  the given transaction itself  $\hat t$, unconfirmed transactions in the mempool $T_{m}$ and information recorded in the blockchain $ChainInfo$. Furthermore, $ChainInfo$ can be divided into two parts, confirmed transactions $T$ recorded in the blocks and the block-related information $B$ in the blockchain. Meanwhile, different estimation models have their specific extra  parameters $Const$. Hence, the transaction fee estimation problem can be formulated as:
\begin{displaymath}
	\hat r = \mathcal F(\hat t, T_{m}, ChainInfo, \{\theta,\vartheta\},Const)
\end{displaymath}
where $\theta$ (block interval) and  $\vartheta$ (time interval) refer to the expected confirmation time, and $\mathcal F$ refers to a model that takes in all kinds of parameters and produces an estimated fee $\hat r \in \mathbb R_{\ge 0}$.

\begin{table*}[htbp]
\centering
\caption{\protect \label{Parameter} Features in different models}
\begin{tabular}{clllcccl}
     \hline
     \multicolumn{2}{c}{\multirow{2}*{Stream}}  &
    \multicolumn{2}{c}{Features}&\multicolumn{4}{c}{Algorithms}\\
    \cline{3-8}
     && Field & Explanation& BtcFlow&MSLP&BCore&FENN\\ \hline
     \multirow{6}*{$\hat{t}$} &
     &$c_{i}$& transaction input count& & & &$\surd$\\
    & &$c_{o}$& transaction output count& & & &$\surd$\\
     &&$ s$& transaction size& & & &$\surd$\\
     &&$w$& transaction weight&$\surd$ &$\surd$ &$\surd$ &$\surd$\\
     &&$h_{e}$& the height of entering the mempool &$\surd$ &$\surd$ &$\surd$ &$\surd$\\
     &&$ v$& a protocol-related transaction version &$$ &$$ &&$\surd$\\
     \hline

      \multicolumn{2}{c}{\multirow{6}{*}{MemInfo ($T_{m})$}}
        &$w$& transaction weight&$\surd$ &$\surd$ & &\\
     &&$r$& transaction feerate&$\surd$ &$\surd$ &$\surd$ &$\surd$\\

    & &$ h_{e}$& the  height of entering the mempool &$\surd$ &$\surd$ &$\surd$ &$\surd$\\
     &&$ h_{l}$& the  height of leaving the mempool& & &$\surd$ &\\
     &&$ h_{c}$& the height of confirmation& &$\surd$ &&$\surd$\\
     &&$ c^{T_{m}} $& the count of mempool transactions&& &$\surd$ &$\surd$\\
     \hline

     \multirow{13}*{ChainInfo} &
     \multirow{7}*{$T$} &
     $c_{i}$& transaction input count& & & &$\surd$\\
     &&$ c_{o}$& transaction output count& & & &$\surd$\\
     &&$ s$& transaction size& & & &$\surd$\\
     &&$ w$& transaction weight&$\surd$&$\surd$ & &$\surd$\\
     &&$ r$& transaction feerate&$\surd$ & $\surd$&$\surd$ &\\
     &&$ f$& transaction fee& & $ $& &$\surd$\\
     &&$ h_{c}$& the block height of confirmation& &$\surd$ &$\surd$&$\surd$\\

     \cline{2-8}
     &\multirow{6}*{$B$} &
     $ q$& time interval to last block& & & &$\surd$\\
     &&$ s^{b}$& block size& & & &$\surd$\\
     &&$  d$& block difficulty& & & &$\surd$\\
     &&$ w^{b}$& the sum of transaction weight in a block& & & &$\surd$\\
     &&$ c^{b}$& the count of transactions in a block& & & &$\surd$\\
     &&$r^{b}$& average feerate in a block & & & &$\surd$\\

\hline
\end{tabular}

\end{table*}

Table \ref{Parameter} summarizes the parameters of the three categories information used in different transactions fee models presented in this paper. In Section \ref{secExperiments}, we will compare the performance of these models  with their respective  features. The main notations used in this paper can be found in Table \ref{notations}.

\begin{table}[H]
\centering
\caption{\protect \label{notations}Main notations}
    \begin{tabular}{ll}
    \hline
    Notation & Description \\ \hline
    $\theta, \vartheta$& The expected confirmation time \\
    &(block interval and  time interval)\\
    $t$   & a transaction instance  \\
    $w$ & transaction weight  \\
    $r$ & transaction feerate  \\
    $r_{min},r_{max}$ & the  minimum and maximum feerate  \\
    $f$ & transaction fee  \\
    $h_{n}$& current block height  \\
    $h_{e}$& the height when entering the mempool   \\
    $h_{l}$& the  height when leaving the mempool  \\
    $h_{c}$& the height when confirmation \\
    $b$   & a block instance  \\
    $T_{m}$ & transactions  in  the mempool  \\
    $T$& confirmed transactions in the blockchain\\
    $B$& blocks recorded in the blockchain\\
    $Const$& model-relative constriction parameters\\

    \hline
    \end{tabular}%
\end{table}%

\section{BtcFlow}
\label{secBtcFlow}

BtcFlow\textsuperscript{\ref{BtcFlow}} infers the transaction fee based on analysing the rate of   transaction entering the mempool  and transaction confirmation leaving the mempool within  a given time interval. Specifically, it models the blockchain mempool as a container with an inflow pouring in (transaction submission) and an outflow flowing out at at a specified rate (transaction confirmation  is assumed following a specific poisson distribution). The ultimate aim is to empty the container within the given time (the mempool is emptied with no unconfirmed transactions left). During the procedure,
the estimated transaction feerate 
serves as a valve to regulate the input flow so that the container can be emptied within the given time.

In terms of data sources, the future inflow is simulated using the confirmed transactions in the Bitcoin blockchain, the outflow is calculated through a specified function, and the container begins with the unconfirmed transactions in the current mempool. Thus, the general estimation procedure of BtcFlow can be formulated as:
\begin{displaymath}
	 \hat r= BtcFlow(\hat t,T_{m},\{T\},\vartheta ,Const_{Btc})
\end{displaymath}
with $Const_{Btc}=\{BLOCK,p,h_{n}\}$, where $h_{n}$ refers to current block height, $p \in [0,1]$ is a  predefined parameter used to set the block production rate, whereas $BLOCK$ specifies the capacity of transactions that can be confirmed in a single block.  Given a $ p$ and a $ BLOCK$, the outflow within the given time interval $\vartheta$ is then calculated.

\subsection{Estimation procedure}

\begin{figure*}
\centering
\includegraphics[width=125 mm,height=5.5cm]{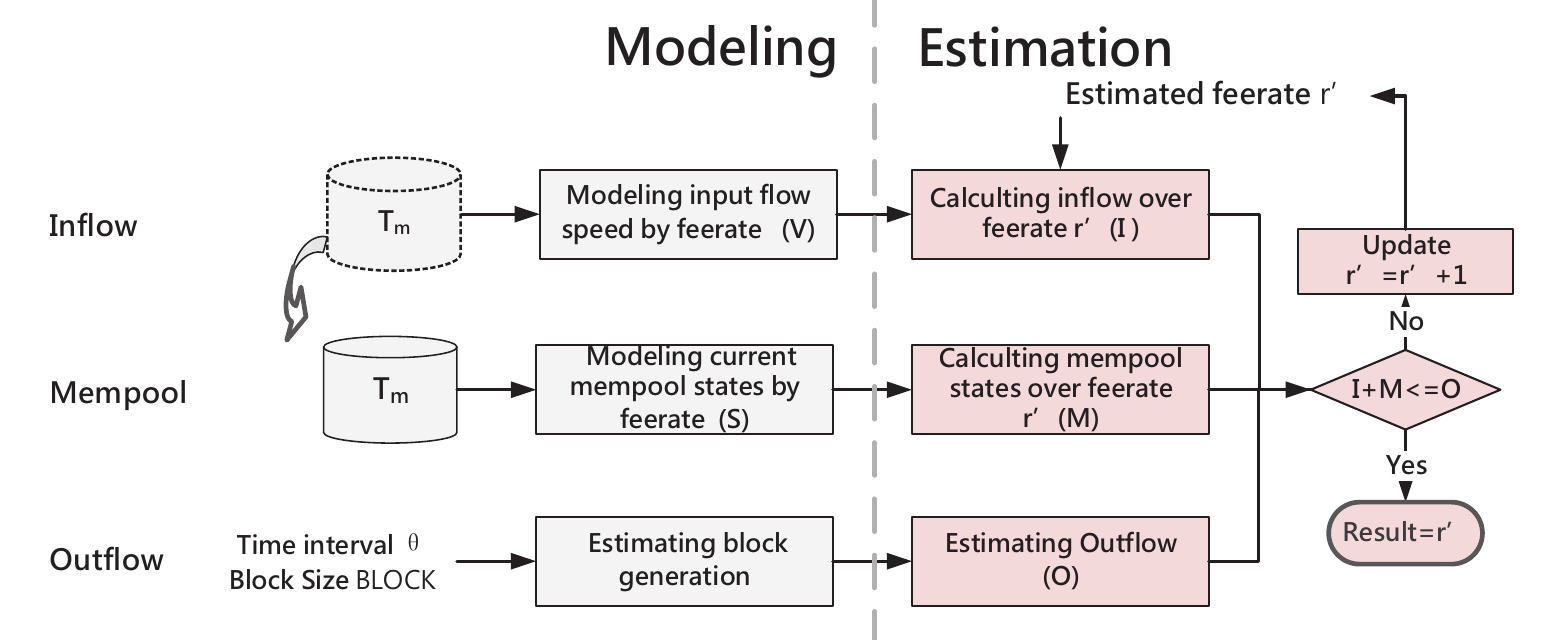}
\caption{\protect\label{BtcFlowfigure}BtcFlow Framework.}
\end{figure*}

BtcFlow works under the assumption that transaction confirmation follows the rules that higher feerate transactions  always have priority to be confirmed earlier than the lower ones. It means that when estimating the transaction fee, BtcFlow only needs to focus on  estimating the minimum boundary feerate transactions which can ensure the unconfirmed transactions with higher feerates ( equal to or larger than the minimum fee rate) can be drained within the given time. The estimating process can be divided into two parts: \emph{The modelling process}, which is designed to analyse the inflow stream, current container state, and the outflow stream of all feerate-range transactions in the given expected time interval, and \emph{The estimation process}, which is used to generate the target transaction feerate based on the simulated data streams. The process is illustrated in Fig. \ref{BtcFlowfigure} (and Algorithm \ref{AlgorithmBtcFlow}).

\begin{itemize}

  \item \textbf{Modeling process}. In this process, BtcFlow  monitor three data streams of transactions in terms of   their transaction feerates: (1) the submitted transactions entering the mempool (the inflow), (2) the unconfirmed transactions in the current mempool (the current container state), and (3) the confirmed transactions leaving the mempool (the outflow). 
  

 Considering that the value of feerate is continuous, it is impractical to compare the transaction flow at each feerate scale. As a result, Btcflow makes a trade-off between the accuracy of its estimations and the complexity of its computations. For simplicity, BtcFlow discretizes the original feerate scales into different integer scales (rounding up the original feerate value to the nearest integer). It means that  BtcFlow models each flow on an integer scale.

\begin{enumerate}
  \item

\emph{The inflow stream} aims to forecast transactions at each feerate scale submitted to the mempool in the future. BtcFlow generates this information from the recent transactions ($T_{Btc}$) entering the mempool over a time interval ($2*\vartheta$). In Btcflow, the speed of transaction submission in terms of transaction size (transaction weight $w$) is used to model the inflow stream, which  can be written as $V=\{v_{u}\}$ where $v_{u}$ is the increasing speed at 
feerate scale $u \in \mathbb{N^{+}}$, 
$u \in [u_{min},r_{max}]$, $u_{min} = 1$ and $u_{max} = 1000$. In fact,  the inflow speed at a specific $u$ is the average speed of transactions with feerates which meet $\lceil r\rceil= u$.
\begin{equation}
\label{inflow}
v_{u}=\sum_{{ t \in T_{Btc}, \lceil r\rceil= u}}\frac{w}{2* \vartheta}
   \end{equation}
\item \emph{Current mempool states} $S=\{s_{u}\}$   describes the volume at each feerate scale in the current mempool $T_{m}$. Similarly, the current mempool state at each feerate scale $u$ is simulated by aggregating weight of transactions in it.
 \begin{equation}
 \label{feer}
s_{u}=\sum_{{t \in T_{m},\lceil r\rceil= u}}w
\end{equation}
\item \emph{The outflow stream}  models the future transaction confirmation over the given expected confirmation time interval. It assumes that a set number of transactions (the size of which is equal to the block capacity $BLOCK$) will be confirmed once a block is generated. As a result, the target of modeling the outflow stream is transferred  to estimate the number of blocks $k$ that will be generated in the specified time frame. BtcFlow assumes $k$ follows the Poisson distribution:
\begin{displaymath}
  P(x=k)=\frac{\lambda^{k}}{k!}e^{-\lambda},k=0,1,2,3,\ldots
\end{displaymath}
 , where $\lambda=\vartheta/10$ (block generation on an average 10-minutes basis). Then, using the parameter $p$, BtcFlow estimates the block count $c_{\vartheta}$ for this time period as follows:
\begin{equation}
\label{counts}
  c_{\vartheta}=Max\{k\mid 1-P(x\leq k)>p\}
\end{equation}
Finally, the transaction outflow can be calculated according to Equation (\ref{outflow}).
\begin{equation}
\label{outflow}
O=c_{\vartheta}*BLOCK
\end{equation}

\end{enumerate}

 \item  \textbf{Estimation process}. The estimation process is to find the minimum feerate to ensure the outflow stream of transactions with higher feerates exceeds (or equals) the input stream and current mempool states of the mempool. BtcFlow's simulation on transaction confirmation in Equation (\ref{outflow}) shows that the outflow confirmation is simply dependent on variables $p$ and $\vartheta$. In other words, the outflow $O$ is constant for a given $p$ and $\vartheta$ as shown in Equation (\ref{outflow}). As Btcflow tests on lower feerate results, the inflow stream and mempool states involve more lower feerate transactions. The inflow and mempool states will eventually exceed the outflow. The feerate next to this border feerate will be returned by BtcFlow.
 
  Basically, the transaction inflow volume $I$ can be calculated as:
      \begin{equation}
       I=\vartheta \sum_{u\ge u'} v_{u}
      \end{equation}, which represents the increase volume in the mempool coming from higher feerates.  The current mempool volume  of higher feerates $M$ is:
      \begin{equation}
      M=\sum_{u\ge u'} s_{u}
      \end{equation}
      . BtcFlow literately tests on the estimated feerate (scale) $u'$ starting from
$u'=u_{max}$  to find the minimum feerate (scale) to meet the condition:
\begin{equation}
   I+M\leq O
\end{equation}
To be specific,  it will start by checking $u'=u_{max}$ first. Only transactions with a feerate scale larger than or equal to $u'$ will be checked. If the outflow exceeds (or equal) the inflow stream and current mempool states, $u'$ will be reduced, implying that the inflow stream and current mempool states will calculate more lower transactions. The estimation procedure will not end until this criteria has not been met, at which point it will return the most recent feerate $u'$ that fulfills the criteria.

\end{itemize}

\begin{algorithm}
  \caption{ BtcFlow algorithm}
  \label{AlgorithmBtcFlow}
   \begin{algorithmic}[1]
       \Require $\hat t$, $T_{m}$, $\vartheta$ and $Const_{BtcFlow}$
   \Ensure
      $ \hat r$
    \State Model the inflow speed $V$ of various feerates in the future $\vartheta$ baded on historical records;

    \State Model current mempool states $S$ of in current mempool $T_{m}$;

    \State  Model the outflow stream $O$ over the future $\vartheta$ according to Equation (\ref{counts}) and (\ref{outflow}) in the future;
     \State Set initiate  feerate  filtering condition $u'$ to the maximum value $u_{max}$;
    \label{coreStart}

   \While{$ u'>0$}
   \State Calculate the inflow volume $I$ of higher feerates  over the future $\vartheta$;
    \State Calculate the current volume of higher feerates $M$ in mempool over $\vartheta$;

    \If {$ I+M> O$}
    \State \textbf{return} $u'$;

    \Else
    \State Update feerate $u'=u'-1$;
    \EndIf
\EndWhile \label{coreEnd}

  \State \textbf{return}  The estimated feerate $\hat r=u'-1$;

  \end{algorithmic}
\end{algorithm}

\subsection{Algorithm analysis}
\label{btcAlgorithmAnalysi}
By simulating newly submitted transactions entering the mempool, transactions in the mempool and transaction confirmation in the blockchain, BtcFlow returns the smallest feerate to ensure that the confirmation exceeds the inflow submission and current transactions in the mempool within the expected time interval. However, this model is faced with several  limitations in its outflow modeling:

\begin{itemize}

\item    It assumes that a size of $BLOCK$ of higher-feerate transactions will be confirmed within a block interval. In fact, in the blockchain system, some smaller feerate transactions can also be confirmed in a block due to some reasons. In addition, the size of a block can be various in the blockchain and $BLOCK$ is the upper bound of a block. 
\item  The possibility parameter $p$ can fail to work. BtcFlow calculates block generation using three different confidence intervals: `optimistic', `standard' and `cautious' with the possibility parameter $p \in\{0.5, 0.8, 0.9\}$. In some cases,  $p$ may not work. According to the result for $\vartheta=10$ in Table 3, the estimated block count $c_{\vartheta}=0 $ under $p=0.8$ and $p=0.9$, which means that $p$ fails to control the accuracy of block generation as effectively as it should be.
\item The assumption of a 10-minute block generation interval may be incorrect. In the Bitcoin blockchain system, the time it takes to generate a block varies.
\end{itemize}

\begin{table}[htbp]
\label{conf}
\centering
\caption{ The estimated  block count $c_{\vartheta}$ generated in the expected time $\vartheta$ (minutes) according to different possibility $p$}
\begin{tabular}{ccc}
   \hline
  $c_{\vartheta}$ & $\vartheta=10 $ & $\vartheta=20 $ \\
  \hline

0 &[0.632,1.000]& [0.864,1.000]\\
1& [0.264,0.632) &[0.594,0.864)\\
2 &[0.080,0.264)& [0.864,0.323)\\
3& [0.019,0.080)& [0.323,0.143)\\
4 &[0.004,0.019)& [0.143,0.053)\\

  \hline
\end{tabular}

\end{table}

\section{Bitcoin Core (BCore)}
\label{secBCore}

 BCore\textsuperscript{\ref{BCore}} is what the community calls the codebase used to control the currency.  It gives an estimate of the transaction feerate for a given block interval. Similar to BtcFlow, BCore seeks out the lowest feerate to ensure that transactions are confirmed within the specified timeframe. The distinction is that BCore pays more attention to historical transactions that have been recorded in the blockchain. It argues that by combining the feerates of prior transactions confirmed within the same block interval, an expected feerate may be determined. Meanwhile, when computing the estimation, BCore considers the records in recent blocks to be more useful than those in previous blocks.

This prediction procedure can be formulated based on its data resources:
\begin{displaymath}
	\hat r= BCore(T_{m},\{T\},\hat t,\theta,Const_{BCore})
\end{displaymath}
 where $Const_{BCore}=\{\alpha,h_{n},p_{1},p_{2}\}$, where $\alpha$
 is a decay parameter $\alpha\in (0,1]$ used to scale the effect of different blocks. The impact of previous blocks will fade by a fraction of $\alpha$ when a new block is mined. $p_{1}$ and $p_{2}$ are used to control the estimation procedure. $p_{1}$  is used to ensure that enough samples have been obtained, while $p_{2}$ is used to manage the percentage of successfully confirmed transactions. For training, BCore defaults to using all historical confirmed transactions prior to the current block height $h_{n}$.

\subsection{Estimation procedure}
\begin{figure*}
\centering
\includegraphics[width=128 mm,height=6cm]{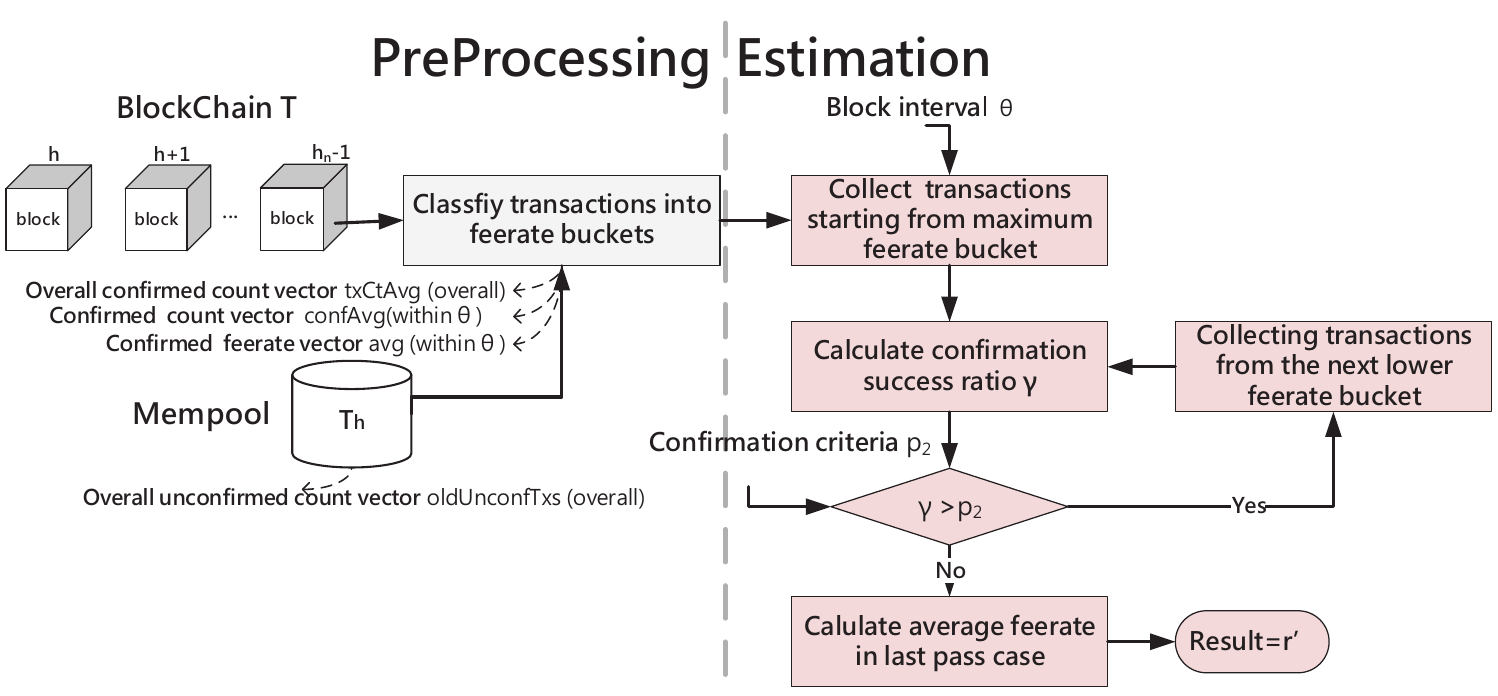}
\caption{\protect\label{BCoreFramwork}BCore framework.}
\end{figure*}

The estimation procedure illustrated in Fig. \ref{BCoreFramwork} (more details in  Algorithm \ref{algorithmBCore}) is composed of two phrases: The \emph{preProcessing process} in BCore is to collect transaction confirmation related to the transaction feerate. To be specific, BCore focuses on three parts of information when given a specific feerate and an expected confirmation time: overall historical confirmation in the entire blockchain, transaction confirmation related to this feerate and unconfirmed transactions that have been waiting longer than the given confirmation time. BCore returns the estimated feerate based on transaction records that meets the model requirements during \emph{the estimation procedure}.

\begin{itemize}

  \item \textbf{PreProcessing process}.  When dealing with the historical confirmation information, BCore believes that recent blocks can provide more accurate information than older blocks. The influence is precisely described as an exponentially weighted moving trend. When a new block is formed, the old blocks' impact fades at a set rate\footnote{In Bitcoin Core prior to version 0.15, for example, the decay rate was set at 0.998.}. Considering that feerate values are continuous, BCore utilizes a feerate technique similar to BtcFlow, a trade-off between estimation precision and processing complexity. It uses the term '\emph{bucket}' to define a feerate scale (starting at 1), and instead of examining each feerate value, it looks at the confirmation information for each feerate scale (bucket).
  
  To be more specific, BCore will collect the following three types of information for each transaction following each feerate bucket $u$ based on current block height $h_{n}$, its conformation block height $h_{c}$, and the block height of it entering the mempool $h_{e}$ for each expected block interval $\theta$ as shown in Fig. \ref{BCore}:

  \begin{itemize}

    \item \emph{Overall confirmed transaction count (txCtAvg)} is a decayed number of transactions  that have confirmed before the current block height $h_{n}$. Meanwhile, the confirmed block interval is no more than $\vartheta_{max}$, the prefixed maximum block interval\footnote{$\vartheta_{max}=24$ in Bitcoin Core prior to version 0.15 and $\vartheta_{max}=1008$ in the later version}.
          \begin{equation}
            \label{txCtAvg}
             txCtAvg(u)=\sum_{{ t \in T,r\in u,h_{c}<h_{e}+\vartheta_{max},h_{c}<h_{n}}}\alpha^{h_{n}-h_{c}}
            \end{equation}
            \item \emph{Confirmed feerate (avg)} is the weighted feerate of all confirmed transactions for each bucket $u$, which is related to txCtAvg.
      \begin{equation}
                \label{avg}
                 avg(u)=\sum_{{ t \in T, r\in u, h_{c}<h_{e}+\vartheta_{max},h_{c}<h_{n}}}r\ast \alpha^{h_{n}-h_{c}}
                \end{equation}
       \item \emph{Confirmed  count (confAvg)}, which is the  number of transactions that have been confirmed within $\theta$ and these transactions have been confirmed before current block height $h_{n}$.
             \begin{equation}
                \label{confAvg}
                 confAvg(u)=\sum_{{ t \in T,r\in u,h_{c}-h_{e}\leq\theta ,h_{c}<h_{n}}}\alpha^{h_{n}-h_{c}}
                \end{equation}

            \item \emph{Overall unconfirmed counts (txUnCt)}, which is the number of transactions that have not been confirmed and have waited more than $\theta$ at the current block height.

                \begin{equation}
                    \label{txUnCt}
                    txUnCt(u)=\sum_{{ t \notin T, h_{n}-h_{e}\geq \theta}}1
                \end{equation}

  \end{itemize}

\item \textbf{Estimation process}. BCore will deliver the smallest feerate that ensures this feerate's confirmation within the given time interval has a greater ratio throughout the whole confirmation time range.
    \begin{equation}
      \gamma=\sum_{u'}\frac{confAvg }{txCtAvg  + txUnCt}
    \end{equation}

    To be specific, BCore will begin collecting transactions from the highest feerate bucket first, then add transactions from lower buckets until enough samples have been obtained. The success ratio of those selected transactions will then be calculated.  If the ratio $\gamma$ is greater than the success confirmation ratio $p_{2}$, BCore will empty the current selected transaction collection and begin collecting sufficient samples from the next buckets (lower feerate bucket) until the ratio fails the test ($p_{2}$ in Algorithm \ref{algorithmBCore}). BCore will examine the feerate in the last pass transaction collection if it fails the test. Finally, BCore will return the average feerate in the bucket $u$, which is where the median transaction count of the selected transaction collection is located.
        \begin{equation}
      r'=\frac{avg(u) }{txCtAvg(u)  }
    \end{equation}

\end{itemize}

\begin{algorithm}[t]
  \caption{ \protect\label{algorithmBCore}BCore algorithm}
  \label{core14}
    \begin{algorithmic}[1]
    \Require $\hat t$, $T_{m}$, $T$, $\theta$ and $Const_{BCore}$

    \Ensure: $\hat{r}$
    \State Collect information of transactions confirmed within $\theta$  and unconfirmed transactions from mempool at $h_{n}$ by \textbf{Equation} (\ref{txCtAvg})-(\ref{txUnCt})
    \State Collect sufficient transactions starting from highest feerate buckets\label{collect}
    \State Calculat the success ratio $\gamma$  based on the selected samples and unconfirmed transactions which have waited for more than $\theta$ \label{ratio}
    \While{  $\gamma$ exceeds the confirmation criteria $p_{2}$}
    \State Remove the selected samples as well as their related buckets
    \State Repeat step \ref{collect}-\ref{ratio} based on transactions in the other buckets
    \State Calculate the average feerate $r'$ of transactions in the bucket  where the median of transaction count sits

    \EndWhile
    \State  \textbf{return} estimated feerate $\hat{r} = r'$;
  \end{algorithmic}
\end{algorithm}

\subsection {Updates in Bitcoin Core post v0.15}
There are four major changes in the algorithm after version 0.15:
\begin{itemize}
  \item The feerate bucket interval is reduced to 5\% from 10\%  to enable more accurate estimation.
  \item It can handle up to 1008 blocks of maximum expectation estimate interval. Meanwhile, the estimation process has become more complex. The estimation process is repeated over three time horizons: short history (targets up to 12 blocks), medium history (targets up to 48 blocks), and long history (targets up to 1008 blocks).
\item It provides two types of feerate estimate results. \emph{Conservative estimation} use longer time horizons to produce an estimate which becomes less sensitive to sudden changes in fee conditions. \emph{Economical estimation} use shorter time horizons and will focus more on the short-term changes in fee conditions.
\item Transactions which leave the mempool due to eviction or other reasons are taken into account by the fee estimation logic  in $txUnCt$.
\end{itemize}

\subsection {Algorithm analysis}
By examining the confirmation behavior of transactions in various feerate scales, BCore calculates the predicted feerate. When it comes to unconfirmed transactions in the mempool, however, it only considers those that have taken longer than expected. It ignores the competition from transactions with a lower wait time or that have just been submitted to the mempool. As a result, BCore may be oblivious of unexpected transaction fee change in the mempool, especially when a significant number of higher feerate transactions are submitted at once, resulting in severe confirmation competition.

\section{ Mempool state and linear perceptron machine learning (MSLP)}
\label{secMSLP}
MSLP \cite{al2018bitcoin} transforms the transaction fee problem into a classic binary classification problem, predicting whether a transaction can be confirmed within a specified timeframe based on its feerate. To be specific, MSLP first calculates an estimated confirmation time based on the mempool.  The estimated confirmation time is then compared to the historical confirmation time in the blockchain. If the estimated confirmation time exceeds the recorded confirmation time, it means that this transaction can be confirmed earlier under this feerate. In other words, this charge assures that the transaction will be completed within the estimated confirmation time. Finally, MSLP uses the capabilities of a neural network to learn the confirmation result (success or failure). It focuses on transaction priority information in the mempool, as opposed to BtcFlow, which focuses on transaction feerate flow change in the mempool.

The prediction procedure can be formulated  based on its data resources:
\begin{displaymath}
	\hat r= MSLP(T_{m},\{T\},\hat t,\theta,Const_{MSLP})
\end{displaymath}
with $Const_{MSLP}=\{BLOCK,SLICE,h_{n}\}$. When MSLP simulates the confirmation time of a transaction in the mempool, the capacity of each block is assumed to be $BLOCK$. Similarly, each block is broken into smaller chunks of $SLICE$ size. By default, transactions confirmed in the block height anterior to the current block height $h_{n}$ are used in the training process.

\subsection{Estimation procedure}
\label{Estimation Procedure}

\begin{figure*}
  \centering
\includegraphics[width=140 mm,height=6.2cm]{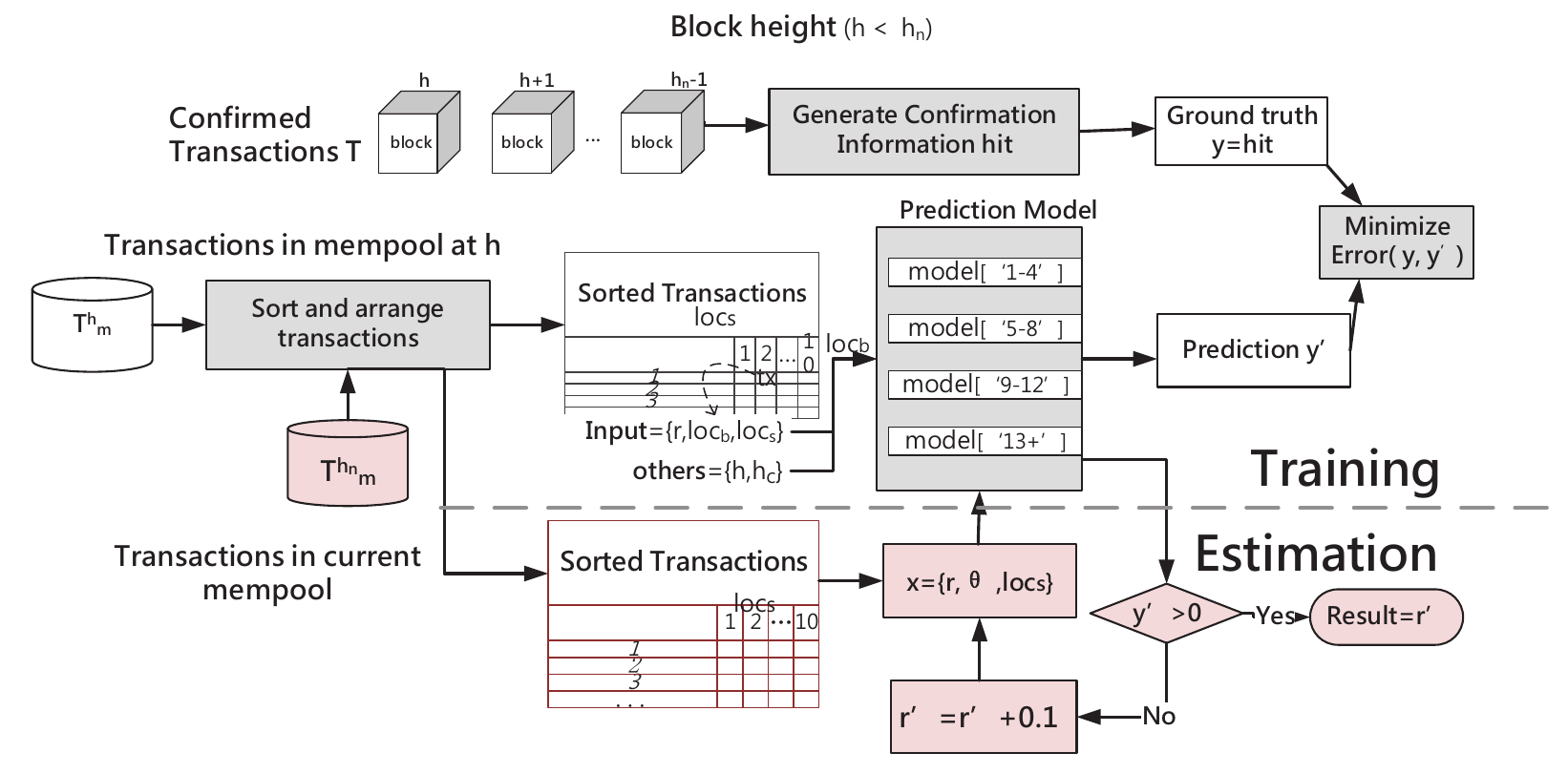}
\caption{\protect\label{MSLP}MSLP framework.}
\end{figure*}

The estimation procedure as shown in Fig. \ref{MSLP} (and Algorithm \ref{mempoolLinear}) is composed of two phases: the training process and the estimation process. \emph{The training process} is used to discover the inherent patterns among transaction confirmation time, current memepool, and the transaction feerate. \emph{In the estimation process},  given the confirmation time, MSLP will return the transaction feerate to ensure confirmation based on the learnt patterns in the training process.

\begin{itemize}
  \item  \textbf{Training process}. In the mempool, unconfirmed transactions compete for confirmation in the blockchain. To highlight this competitiveness, MSLP coined the term `expectation confirmation time' to represent the transaction processing priority. The unconfirmed transaction is treated by MLSP as a queue sorted by transaction feerate, and only a block of transactions (related to the fixed capacity of a block in the Bitcoin blockchain system) in this queue can be confirmed at a time. As a result, the expected confirmation time of a transaction is proportional to the number of blocks in the queue before it is processed. Meanwhile, certain other unseen elements may have an impact on the confirmation procedure. By training the model with history confirmation instances, MSLP leaves the analysis of the intricate link between the expectation confirmation time, blockchain confirmation time, and transaction feerate to a neural network.

      \begin{enumerate}

        \item \emph{Estimating expectation confirmation time}.  The sorted unconfirmed transaction queue is divided into a series of virtual blocks with a size of $BLOCK$ by MSLP. Virtual blocks with higher transaction fees are expected to be confirmed sooner than those with lower transaction fees. Meanwhile, each block's sorted transactions are separated into sequence slices with the size $SLICE$. Finally, the virtual block sequence position $loc_{b}$ and the slice sequence position $loc_{s}$  combine to form the expectation confirmation time  for a transaction $t$ with feerate $r'$ as shown in Equation (\ref{blockNum})-(\ref{sliceNum}): 
          \begin{equation}
            \label{blockNum}
             loc_{b}=\sum_{t \in T^{h}_{m},h<h_{n},r\geq r'} w/BLOCK+1
            \end{equation}
           \begin{equation}
            \label{sliceNum}
             loc_{s}=\sum_{t \in T^{h}_{m},h<h_{n},r\geq r'} w/SLICE+1
            \end{equation}
. $T^{h}_{m}$ is the transaction collection in the mempool at block height $h_{n}$.
          \item \emph{Constructing training instances}. In MSLP, the inputs for instances are the feerate $r'$ and the expectation time information ($loc_{b}$ and $loc_{s}$), and the output reflects whether this transaction can be confirmed within its expectation confirmation time. The model output will be $hit=1$ if the expectation virtual block position $loc_{b}$ is larger than the actual confirmation time (the interval between the research block height and final confirmation block height), indicating that it was confirmed earlier than expected. Otherwise, the model output would be $hit=0$.

             \begin{equation}
                hit=
                \begin{cases}
                    1& loc_{b}\geq h_{c}-h\\
                    0& loc_{b}< h_{c}-h
                \end{cases}
            \end{equation}
              In fact, an unconfirmed transaction  can  generate several failure instances with $hit=0$ before it is finally confirmed in the blockchain.

           \item \emph{Training models}.
           To deal with varying expectation block intervals, MSLP employs four training models: `[1-4]',`[5-8]',`[9-12]' and `[13+]'. For example, the instance $x=[r,loc_{b}=5,loc_{c}]$ will be applied to train the '[5-8]' model. Each model shares the same network configurations by applying  a neural network layer with a linear activation function to generate the classification result.
        \end{enumerate}

  \item \textbf{Estimation process}. In this process, MSLP returns the estimated feerate related to the given expected interval $\theta$ based on the trained models.

  MSLP first constructs a pseudo input instance $x=[r',\theta,loc'_{s}]$ based on current mempool, where $r'$ is the initial feerate  along with a slice position $loc'_{s}$. When constructing this instance, two cases arise related to the given block interval:
      \begin{description}
        \item[Case 1] Out of boundary error. It occurs when the anticipated interval $\theta$ surpasses the current mempool's maximum virtual block position. MSLP will throw an error for input invalidation in this situation.

        \item[Case 2] Setting the slice position in  $x=[r',\theta,loc_{s}]$. $loc_{s}$  is usually the last slice in virtual block $\theta$, and $r'$ is the smallest feerate in this slice. However, it is possible that the virtual block at position $\theta$ is not full, and so the smallest feerate in the last slice cannot be provided. In this scenario, $x=[r'=0,\theta,loc_{s}]$ and $loc_{s}$ is the actual virtual slice position in the virtual block.
      \end{description}

      According to $\theta$, after the testing instance is built, it will be transferred to the specific trained model. The feerate in this case is the estimated result from MSLP if the model returns $hit=1$. Otherwise, MSLP will incrementally update feerate $r'$ in the testing instance until it returns $hit=1$.

\end{itemize}

\begin{algorithm}[t]
  \caption{ MSLP algorithm}
  \label{mempoolLinear}

     \begin{algorithmic}[1]
       \Require $\hat t$, $T_{m}$, $T$, $\theta$ and $Const_{MSLP}$
    \Ensure:$ \hat r$
    \State Collect and sort transactions  in the mempool at each block height $h<h_{n}$. \label{1}
    \State Construct training instances including the feerate, the expectation confirmation time, and the confirmation information $hit$ for each transaction.\label{2}
    \State Train different neural network models with instances
    \Comment  The estimation process starts
    \If{ $\theta$ exceed the maximum virtual block position in current mempool $h_{n}$}
    \State Return an invalid input error  \Comment  Case 1
    \Else \Comment  Case 2

    \State Construct the testing instance $x=[r',\theta,loc'_{s}]$
    \State  Predict confirmation result for $x$ \label{retrain1}
    \While { Prediction result returns FALSE} \Comment  TRUE if activation$\geq 0$ else FALSE
      \State  Update $x$ with $x=[r'+0.1,\theta,loc_{s}]$
      \State  Predict confirmation result for the new testing instance
    \EndWhile
    \State  Estimated feerate $\hat r=r'$ \label{retrain2}
     \EndIf
   \State  \textbf{return}  $ \hat r$;

  \end{algorithmic}
\end{algorithm}
\subsection{Algorithm analysis}
MSLP achieves feerate estimation based on analysing the inherent patterns among the  confirmation time, transaction ranking in the mempool, and the transaction feerate. Unfortunately, it fails to deliver an estimation result when the confirmation time exceeds the expected confirmation time, or when there are no previous training instances available for the associate training models.

These instances are common throughout the block generation process. Meanwhile, as the training process demonstrates, there is no clear differentiation between different block intervals. For example, cases with a block interval of 5 to 8 are used to train the same model `[5-8]'. Another potential constraint is the use of block capacity in the assessment of expected confirmation time. Except for the geniue reward transaction from the Bitcoin blockchain system, a block in the blockchain can be empty with no unconfirmed transactions.

\section{Fee estimation based on neural network (FENN)}
\label{secFENN}
Due to the low block capacity, the majority of submitted transactions may experience various confirmation delays. Transactions are selected and added to the miner's mempool after submission, where they compete for confirmation in the next block. A transaction is considered complete when it is recorded in a block in the blockchain. In the confirmation process, transaction fees are considered as an incentive to confirm transactions into the blockchain. To sum up, we summarize three groups of features that may influence the transaction confirmation:
 
  \begin{itemize}
     \item \textbf{Transaction features}, which describes the submitted transaction.
     \item \textbf{Mempool states}, which records the distribution of feerates of unconfirmed transactions in the mempool, implicitly modelling the competition among unconfirmed transactions.
      \item \textbf{Network features}, which reflects the characteristics of the mined blocks including block size, block generation speed, etc.

  \end{itemize}

  These three groups of features correspond to the three types of information fed to the estimation function $\mathcal{F}$ in Section~\ref{sectProblemDefinition}. Although transaction features are already available in the submitted transaction, network features and mempool states are not known. However, such features are desirable, because if we had known how many transactions would be contained in future blocks, how fast future blocks would be generated, how competitive the submitted transaction would be in future mempools, we would increase the chance to predict the confirmation fee more accurately. Consequently, in FENN, our main idea is to predict network features and mempool states from historical state sequences by utilizing sequence learning models. Finally, we combine the three groups of features to do the estimation.

The prediction procedure can be formulated  based on its data resources:
\begin{displaymath}
	\hat f= FENN(T_{m},\{T,B\},\hat t,\theta,Const_{FENN})
\end{displaymath}
with $Const_{FENN}=\{h_{n}\}$. $\theta$ is the expected confirmation time for a transaction. In the estimation, transactions confirmed previous to current block height $h_{n}$ are used in the training process.

\subsection{Estimation procedure}

\begin{figure*}
\centering
\includegraphics[width=125 mm,height=5cm]{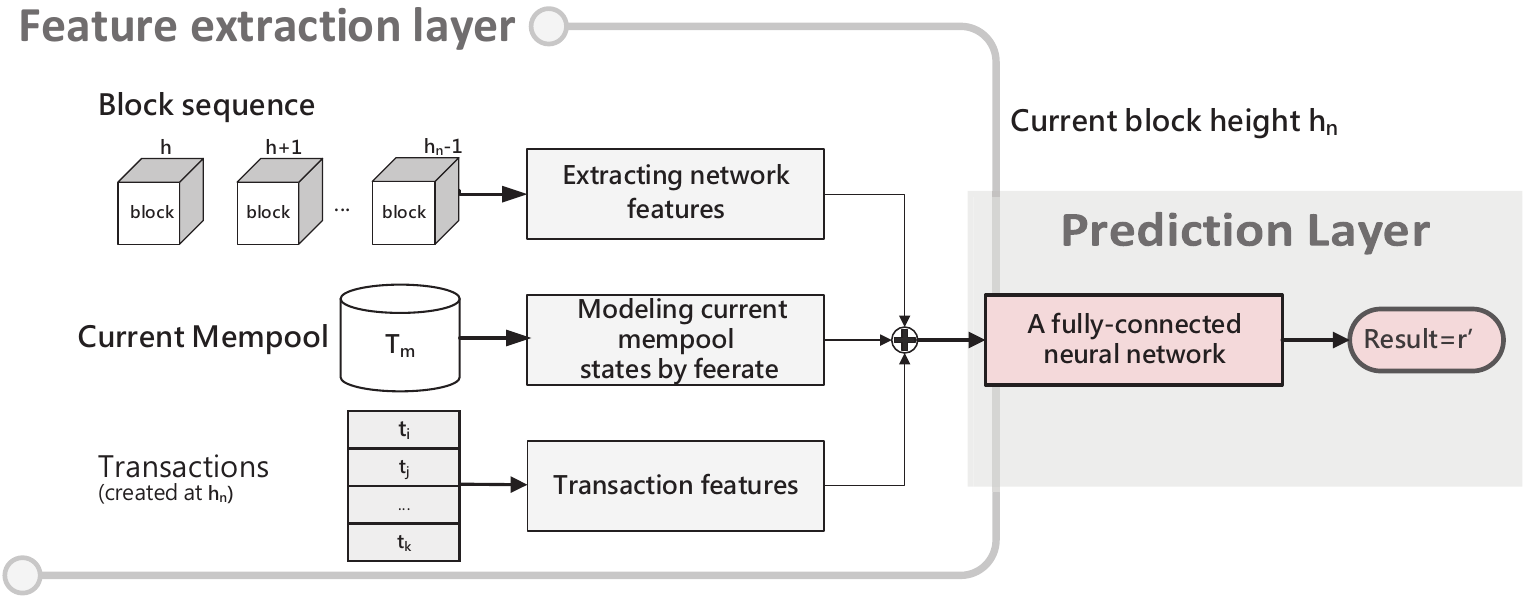}
\caption{\protect\label{FEN}FENN framework.}
\end{figure*}

 The estimation framework can be divided into two layers, one \textit{feature extraction layer} to extract patterns from network features, mempool states and  the submitted transaction itself, and one \textit{prediction layer} to analyze the relationship between transaction fee and the extracted features. Fig.~\ref{FEN} shows the framework.
\subsubsection{Feature extraction layer}
It includes three parts. Other than modelling the submitted transaction itself, the feature extraction layer also predicts the future characteristics of block states and model mempool competition states of the unconfirmed transactions.

    \begin{enumerate}
    
       \item  \emph{Transaction features}  contain information on the transaction that has been submitted and is awaiting confirmation. We pick features that we believe may affect a transaction's validation and confirmation. The transaction vector contains:
       \begin{itemize}
         \item  \emph{number of inputs}, \emph{number of outputs} Miners need to seek for the source transactions pointed to the new transaction's inputs when confirming a transaction, which means that the number of transaction inputs and outputs affects the verification complexity.

        \item \emph{transaction version}, \emph{transaction size and weight} We use both transaction size of raw data and transaction weight to characterize transactions.
      \item    \emph{transaction first seen time}, \emph{confirmation timestamp} and \emph{confirmation block height}. The first-seen time refers to the time that a transaction is first observed. Because it's difficult to determine the precise submission time of a historical transaction, we use the publicly available first-seen time.
       \end{itemize}

      \item \emph{Current mempool states} indicate the transactions feerate distribution in the mempool. Inspired by `bucket' term in BCore, we classify our transactions in the mempool into different feerate buckets  and make use of the transaction count in each bucket $u$ to model mempool states $T_{m}$. The bucket interval is the same as BCore with 5\% increasing interval for accuracy.
         \begin{equation}
             \label{memLstm}
            s_{u}=\sum_{{\substack{ t \in T_{m},\\r\in u}}}1
        \end{equation}

      \item \emph{Network features}  are expected to encode future block size and generation speed, which can affect a transaction's confirmation time. Historical network features are learned as a sequence to predict future network features.
    \begin{itemize}
      \item \emph{block size}, \emph{block weight} and \emph{transaction count} We  use three factors to characterise the size of a block, namely, the overall size of transactions (Bytes), the overall weight of transactions (Weight) and the transaction count in a block.
  
      \item \emph{difficulty}  It reflects the mining difficulty in the Bitcoin system, which is tuned to maintain an average 10-minute block frequency.
      \item  \emph{block time} The mining time of this block. It reveals the block generation speed.
      \item \emph{average feerate in block} The average feerate of  all the transactions in the block. This indicator is designed to reveal the feerate trend in continuous blocks.

    \end{itemize}

  In FENN, future network features are learnt by sequence models \cite{sundermeyer2012lstm,mcnally2018predicting,fu2016using,gers2002applying,zhu2017deep,srivastava2015unsupervised}. We implement two alternatives, LSTM and Attention. In both models, we use ${x}_{i}$ to represent the input feature. To be specific, $x_i$ refers to the network feature $b_h$ at a specific block height and $t$ is the length of the sequence.
  
    \begin{description}
                   \item[Approach 1:] \textbf{LSTM} \cite{hochreiter1997long} extract patterns by aggregating information on a token-by-token basis in a sequential order and summarizes the sequence into a context vector. To be specific,  at each time step, LSTM maintains a hidden vector $h$ and a memory vector $c$ responsible for state updates and output prediction \cite{karim2017lstm}, and the final state is used as the extracted patterns from the sequence in our models:
                        \begin{equation}\label{LSTMEquation}
                        \begin{split}                      i_{t}&=\sigma(W^{i}x_{t}+M^{i}h_{t-1})\\
                        f_{t}&=\sigma(W^{f}x_{t}+M^{f}h_{t-1})\\
                        o_{t}&=\sigma(W^{o}x_{t}+M^{o}h_{t-1})\\
                        \tilde{c}_{t}&=tanh(W^{c}x_{t}+M^{c}h_{t-1})\\
                        c_{t}&=i_{t}\odot \tilde{c}_{t}+f_{t}\odot c_{t-1}\\
                        h_{t}&=o_{t}\odot tanh(c_{t-1})
                        \end{split}
                        \end{equation}
where $[W^{i}, W^{f}, W^{o},W^{c}, M^{i}, M^{f}, M^{o},M^{c}]$ are weight matrices, $x_{t}$ is the vector input to the timestamp t, $h_{t}$ is the current exposed hidden state, $c_{t}$ is the memory cell state, and $\odot$ is element-wise multiplication.
                    \item[Approach 2:] \textbf{Attention} is another popular time-series processing technique. It simulates the cognitive process of selectively concentration on different parts in psychology. In other words, it returns a new representation vector related to the importance at various positions in the sequence. Three state-of-the-art attention modules are applied below:
                        \begin{enumerate}
                          \item  Additive attention \cite{bahdanau2014neural}  computes the compatibility function using a feed-forward network with a single hidden layer.
                        \begin{equation}\label{AdvEquation}
                        \begin{split}
                        &h_{t,t'}=tanh(W^{t}x_{t}+W^{x}x_{t^{'}})\\
                        &e_{t,t^{'}}=\sigma(W^{a}h_{t,t^{'}})\\
                        &a_{t}=softmax(e_{t})\\
                        &x^{'}_{t}=\sum_{t'}a_{t,t'}x_{t'}\\
                        \end{split}
                        \end{equation}
                                                where $[W^{t}, W^{x}, W^{a}]$ are weight matrices, $x_{t}$ is the input token and $x^{'}_{t}$ is the output.
                          \item Self-attention \cite{vaswani2017attention} projects the input sequence X into three different spaces: Q(Query), K(Key) and V(Value). For each token $x_{i}$, it calculates the attention value between the selected token and the other tokens .
                          
                          \begin{equation}\label{SelfEquation}
                        \begin{split}
                        Q&=XW^{Q}\\
                        K&=XW^{K}\\
                        V&=XW^{V}\\
                        X^{'}&=softmax(\frac{Q K^T}{\sqrt{d_k}})V
                        \end{split}
                        \end{equation}
                        where $[W^{Q}, W^{K}, W^{V}]$ are weight matrices, X is the input sequence and $X^{'}$ is the output sequence.

                          \item  A simple weighted attention \cite{felbo2017using} is applied to learn the hidden states at each timestamp in the LSTM layer.
                          \begin{equation}\label{WhtEquation}
                        \begin{split}
                             &a_{t} = softmax(Wh_{t})\\
                             &h^{'}_{t}=\sum_{t'}a_{t,t'}h_{t'}
                                                   \end{split}
                        \end{equation}
                        \end{enumerate}
                        where $W$ is a weight matric, and h is the hidden states in the former LSTM processing stage.
    \end{description}

    \end{enumerate}

\subsubsection{Prediction layer}
After aggregating inputs from the feature extraction layer,  FENN is followed by a  fully-connected neural network. By learning the relationship among historical block information, mempool data, and transaction details, FENN can provide a specific estimated feerate for each transaction. The testing instance of the estimated transaction consists of three parts: the block sequence, current mempool states and the transaction itself.

\begin{algorithm}[t]
  \caption{ FENN Framework Algorithm}
  \label{FENNLSTM}

      \begin{algorithmic}[1]
      \Require $\hat t$, $T_{m}$, $T$, $B$, $\theta$ and $Const_{FENN}$

    \Ensure: $\hat{f}$

    \While{$ h<h_{n}$}
       \Comment  Training process

    \State Extracting network features in the block sequence
    \State Modeling mempool states $S$  in the mempool at block height $h$
    \State Extracting features for transactions submitted during block $h-1$ and $h$
    \State Training model and update $h=h+1$
    \EndWhile \Comment  Estimation process

    \State Constructing testing instance based on current mempool states, block sequence and transaction details
    \State Using the trained model to predict the testing instance

    \State \textbf{return} the estimated fee $\hat{f}$;
  \end{algorithmic}
\end{algorithm}

\section{Experiments}
\label{secExperiments}
The datasets, experimental evaluation metrics, and transaction fee estimation solutions are all introduced in this part. Following that, we run a performance analysis on the experimental data.

\subsection{Experiment settings}

%
\subsubsection{ Datasets and implementation}
We constructed datasets by picking 6 different block intervals at random via Blockchain Explorer\footnote{https://www.blockchain.com/explorer}.  Each dataset has 225 blocks, the first 180 blocks are used for training (about 400,000 transaction instances) and the last 45 blocks for testing (see Table \ref{datasets}). In terms of implementation, the hidden units in the sequence processing module in the feature extraction layer are set to 64, and the sequence length is set to 3. FENN's prediction layer is a fully linked three-layer neural network with hidden units 64, 8 and 1, respectively. The Adam optimizer is used to optimize parameters using stochastic gradient descent (SGD) with a batch size of 1000 while training models. All of the algorithms are written in TensorFlow, and all of the tests are run on a single NVIDIA P100 12GB PCIe GPU.

\begin{table}[ht]
    \centering
		\centering
		\caption{Testing dataset information}
        \begin{tabular}{cllc}
            \hline
            \multirow{2}[0]{*}{Dataset} & \multirow{2}[0]{*}{Block Range} & \multicolumn{2}{c}{Block Time} \\
          &       & mean  & variance \\
            
            \hline
            S1     & 621293-621337 & 475   & 1.16e5\\
            S2     & 621301-621345& 579   & 3.20e5 \\
            S3     & 621310-621354 & 607   & 4.20e5\\
            S4     & 621853-621897 & 801   & 5.20e5\\
            S5     & 622214-622258 & 1171  & 1.34e6\\
            S6 & 622301-622345 & 891   & 7.53e5\\
              \hline
         \end{tabular}%
        \label{datasets}
\end{table}

\subsection {Evaluation strategies}
During test,  RMSE  and MAPE are calculated to evaluate the predictive error. Higher feerate transactions tend to confirm earlier than lower feerate transactions, hence in the fee estimate problem, the lower bound fee is usually returned.  Compared to MAPE, RMSE concentrates more on avoiding high abnormal values, i.e. abnormal transaction fee values, which are outliers, have high impact on the error values.
 \begin{itemize}
        \item RMSE
            \begin{displaymath}
                 \mathit{RMSE}=\sqrt{\frac{1}{n}\sum_{i=1}^n(y_{i}-\hat{y_{i}})^2}
            \end{displaymath}

         \item MAPE 
            \begin{displaymath}
                \mathit{MAPE}=\frac{100\%}{n}\sum_{i=1}^n\mid\frac{y_{i}-\hat{y_{i}}}{y_{i}}\mid
            \end{displaymath}
            where $y_{i}$ and $\hat{y_{i}}$ are the ground truth and the prediction for the test sample $i$, respectively.
        \end{itemize}

 Because of the SegWit upgrade in Bitcoin, a vByte was created to signify transaction size. It is roughly equivalent to four weight units. Typically, transaction feerate are expressed in sats/vByte. As a result,  models with predicted feerates need to be converted to transaction fees using Eq.\ref{feerateequation}. The transaction feein BtcFlow is the integer component of the value according to its official documents.
 
\begin{equation}
\label{feerateequation}
f=\frac{w}{4}*r
\end{equation}

\subsubsection{ Compared methods}
  \begin{itemize}
   \item \textbf{BCore}: We use the latest configuration in BCore (which is the same in V0.15 - V0.21), with a bucket interval of 5\% and three alternative block period modes. 
    \item \textbf{BtcFlow}: In order to simulate block generation speed, BtcFlow offers three distinct probability parameters: `Optimistic', `Standard', and 'Cautious'. The 'Standard' mode is selected, with $p=0.8$.
    
    \item \textbf{MSLP}:  It is a one-layer neural network with a linear activation function.

    \item \textbf{FENN variants}: It includes LSTM models, attention models, and variants with various feature compositions.
        \begin{itemize}
          \item \textbf{LSTM mechanism}: \textbf{LSTM} in Eq. \ref{LSTMEquation}.
          \item \textbf{Attention mechanism}:
            \begin{itemize}
                  \item \textbf{Adv}: Additive attention in ref{AdvEquation}
                  \item \textbf{Self}: Self-attention  in Eq. \ref{SelfEquation}
                  \item \textbf{Wht}: A combined LSTM and a simple weighted attention in Eq. \ref{WhtEquation}
                  \item \textbf{LSTMadv}: A combined LSTM and additive attention
             \end{itemize}
        \item \textbf{Feature compositions on Adv}:
            \begin{itemize}
                  \item \textbf{Adv\_Tx}: Transaction features only
                  \item \textbf{Adv\_BloTx}: Transaction features and network features
                  \item \textbf{Adv\_MemTx}: Transaction features and current mempool states
  
             \end{itemize}

                \end{itemize}

  \end{itemize}

\subsection{Result analysis}
We test on the genuine data to demonstrate the effectiveness and efficiency of FENN transaction fee estimation solution.

\subsubsection{ Estimation results comparison}

\begin{table*}[htbp]
\centering
  \caption{Evaluation of methods  on RMSE }

    \begin{tabular}{llllcllll}
            \hline
    \multicolumn{1}{c}{\multirow{2}[4]{*}{Data}} & \multicolumn{3}{c}{\textbf{Existing work}} & \multicolumn{5}{|c}{\textbf{FENN framework}} \\
\cline{2-9}          & \multicolumn{1}{l}{BCore} & \multicolumn{1}{l}{MSLP} & \multicolumn{1}{l}{BtcFlow}  & \multicolumn{1}{|l}{LSTM} & \multicolumn{1}{l}{Adv} & \multicolumn{1}{l}{Wht} & \multicolumn{1}{l}{Self} & \multicolumn{1}{l}{LSTMadv} \\
    \hline
     S1    & 3.78e4 & 5.00e4 & 3.64e6&  3.63e4 & \textbf{3.46e4} & 3.48e4 & 4.44e4 & 3.54e4 \\
    S2    & 5.23e4 & 5.60e4 & 4.96e6& 4.45e4 & \textbf{4.43e4} & 4.49e4 & 5.70e4 & 4.52e4\\

    S3    & 6.71e4 & 5.56e4  & 5.85e6 &  \textbf{4.64e4} & 4.84e4 & 4.83e4 & 6.84e4 & 4.82e4 \\
    S4    & 5.92e4& 5.95e4 & 1.32e6&  5.13e4 & \textbf{5.00e4} & 5.32e4 & 5.29e4 & 5.25e4 \\
        S5    & 1.05e5& 4.45e4 & 1.76e6  &  3.29e4 & \textbf{3.19e4} & 3.24e4 & 3.20e4 & 3.40e4 \\
      S6    & 4.62e4& 4.79e4 & 2.38e6 &  4.20e4 & \textbf{4.03e4} & 4.11e4 & 4.07e4 & 4.28e4 \\

    \hline
    \end{tabular}%
  
  \label{RMSEError}%
\end{table*}%

\begin{table*}[htbp]
\centering
  \caption{Evaluation of models on MAPE}
    \begin{tabular}{llllcllll}
        \hline
    \multicolumn{1}{c}{\multirow{2}[4]{*}{Data}} & \multicolumn{3}{c}{\textbf{Existing work}} & \multicolumn{5}{|c}{\textbf{FENN framework}} \\
\cline{2-9}          & \multicolumn{1}{l}{BCore} & \multicolumn{1}{l}{MSLP} & \multicolumn{1}{l}{BtcFlow}  & \multicolumn{1}{|l}{LSTM} & \multicolumn{1}{l}{Adv} & \multicolumn{1}{l}{Wht} & \multicolumn{1}{l}{Self} & \multicolumn{1}{l}{LSTMadv} \\
    \hline
    S1 & 432.94 & 140.17 & 19823.93  & 80.79 & \textbf{79.69} & 80.01 & 81.77 & 80.12 \\
    S2& 357.17 & 119.93 & 19882.29  & 78.07 & 79.22 & 79.22 & 91.47 & \textbf{76.23} \\
    S3 & 288.38 & 110.52  & 20270.18  & 75.57 & 75.62 & 75.56& 88.46 & \textbf{71.84 }\\
    S4 & 226.41 & 200.37  & 10392.77  & 42.82 & 41.34 & 44.75 & \textbf{22.55} & 43.86 \\
    S5 & 241.05 & 162.53 & 4070.52  & 64.14 & \textbf{31.84} & 66.55 & 32.61 & 52.31 \\
    S6 & 206.99 & 175.73 & 5468.83 & 59.56 & \textbf{32.94} & 48.87 & 34.36 & 41.06 \\
    \hline
    \end{tabular}%
  \label{EstimationResult}%
\end{table*}%



  

Table \ref{EstimationResult} and Table \ref{RMSEError} show an overall evaluation of performance over various confirmation time.  FENN variants outperform earlier work across all datasets evaluated by RMSE and MAPE. Meanwhile, the models using the additive attention mechanism, Adv and LSTMadv, outperform other FENN models evaluated by MAPE.  Furthermore, Adv has the best RMSE performance for all of the accessible datesets according to Table \ref{RMSEError}.  In other words,  Adv outperforms the other models when it comes to dealing with this estimation problem.

Besides, previous work models perform poorly, with BtcFlow being the worst of them all. Table \ref{valueResult}  demonstrates that each existing model has a significantly higher estimation feerate than the lowest confirmed feerate and the median feerate in the target block, contradicting its feerate processing 
contradicts its assumption of strictly feerate processing priority. In the following section, we will study the effectiveness of our feature framework in FENN.

    \begin{table}[htbp]
    \centering
    \caption{ \protect\label{valueResult} Estimation feerate results from existing work(  $\theta=2$ blocks) }

        \begin{tabular}{cccclcc}
         \hline
             Height & BtcFlow & MSLP  & BCore & Min &Median\\
          \hline
        621339 &  35 & 23.31 & 25.54 & 2.54&21.09 \\
        621340 & 30& 15.93 & 25.54 & 3.0 &21.10\\
        621341 &27 & 23.31 & 25.54 & 1.31&21.16 \\
        621342  &57 & 36.71 & 95.24 & 6.75&21.18 \\
        621343  &42 & 30.23 & 47.85 & 3.01&21.18 \\
      \hline
        \end{tabular}%
    \end{table}%

\subsubsection{Impact of different features in Adv}
\begin{figure}[htbp]
    \centering
    \includegraphics[height=3.4cm]{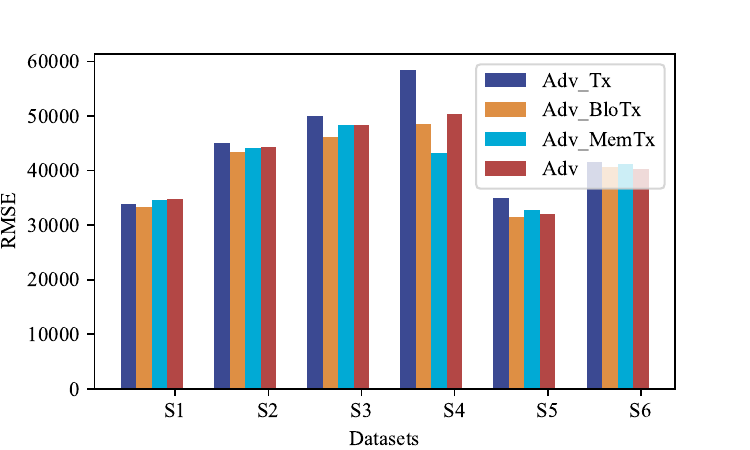}
    \caption{\protect\label{FeaRMSE}Evaluation of feature compositions on RMSE}
  \end{figure}

\begin{figure}[t]
    \centering
    \includegraphics[height=3.4cm]{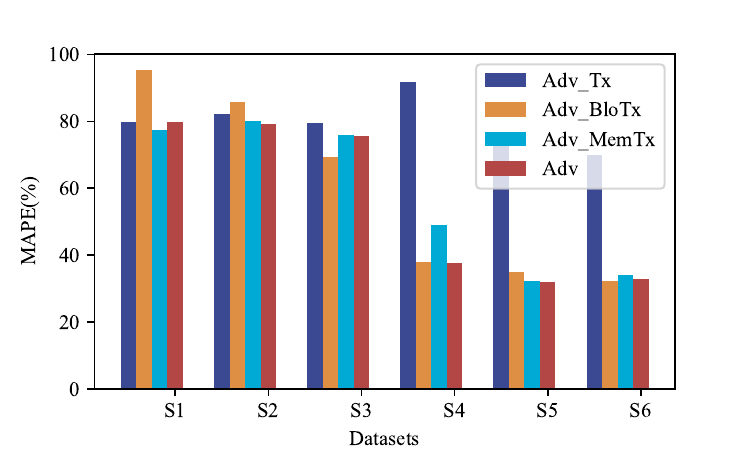}
    \caption{\protect\label{FeaMAPE}Evaluation of feature compositions on MAPE}
		
\end{figure}

We examine four different feature compositions (Adv\_Tx, Adv\_MemTx, Adv\_BloTx, and Adv) in the FENN framework to establish the efficiency of our feature composition. According to Fig.\ref{FeaRMSE} and Fig.\ref{FeaMAPE}, the FENN framework's Adv\_Tx has the poorest performance,  and the accuracy can be improved by introducing mempool states and network features. Specifically, The accuracy of model Adv\_MemTx is increased when mempool states are incorporated into the Adv\_Tx feature structure, as measured by RMSE and MAPE. 

\begin{figure}[t]
    \centering
    \includegraphics[ height=3.4cm]{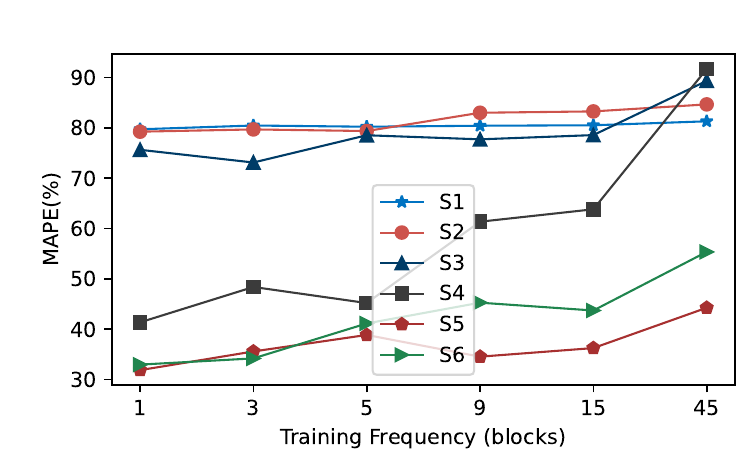}
    \caption{\protect\label{MAPE}Evaluation of  training frequency on MAPE}
\end{figure}

Meanwhile, a same conclusion concerning the effectiveness of network features can be drawn based on the superiority of Adv\_BloTx to Adv\_Tx under RMSE, which is due to its ability to capture blockchain network trends. While network features exhibit a variety of effects evaluated by MAPE, as seen in Fig. \ref{MAPE}. For example, when the block time varies substantially on the datasets S4, S5, and S6, Adv\_BloTx can improve Adv\_Tx's accuracy by approximately 100\%. While network features can have a negative impact on MAPE on S1 and S2 with a steady block time, these issues can be addressed by introducing mempool states, as demonstrated in model Adv. Furthermore, network features can have a modest favorable effect on Adv\_ MemTx when compared to Adv performance on RMSE and MAPE, with the exception of one occurrence on S4 under RMSE. In conclusion, the FENN algorithm benefits from both mempool states and network features, and combining the two parts results in stable outperformance for Adv.

Finally, we compare Adv\_Tx against MSLP, which has already been proved to be the most effective in the existing work in Table \ref{EstimationResult} and Table \ref{RMSEError}. The effectiveness of introducing transaction details in this transaction fee estimate issue is demonstrated by the superiority of Adv\_Tx. In conclusion, FENN demonstrates the effectiveness of introducing transaction features, network features, and mempool states.

\subsubsection{Time efficiency of FENN variants}

 We conduct  experiments to illustrate the efficiency of our proposed FENN framework algorithms. Table \ref{RunningTime} indicates that all FENN variations can complete the training process in one block interval, demonstrating that our framework can handle continuous Bitcoin blockchain data for model updates. Moreover, compared to LSTM-embedded algorithms, the training time for  Adv and $Self$ can be reduced almost 50\%. 

\begin{table}[htbp]
  \centering
  \caption{Training time of FENN framework algorithms with 100 epoches}
    \begin{tabular}{lccccc}
    \hline
          & LSTM & Adv & Wht & Self& LSTMAdv  \\
    \hline
    Training time & 296   & 169    & 323 & 178  & 340   \\
    \hline
    \end{tabular}
  \label{RunningTime}
\end{table}

\subsubsection{Training frequency in Adv}
\label{subsubsec:trainingfrequency}
In prior experiments, Adv has proven to be useful and efficient. Another essential characteristic of Adv is the ability to adapt to new information. We undertake a set of tests to see study its performance with different update frequencies. In our research, we present six different update policies (namely, 1,3,5,9,15, and 45), which imply retraining models at different block intervals. Fig. \ref{FeaRMSE} and Fig. \ref{FeaMAPE} show how Adv performs in terms of accuracy. As we can see, the accuracy of Adv falls as the updating block interval grows. The best frequency policy is one block. Furthermore, when we compare the 3-block technique to the existing work (BCore, MSLP and BtcFlow), we discover that it still outperforms them, implying that our FENN has the ability to incorporate more details in future work.


  \begin{figure}[t]
    \centering
    \includegraphics[ height=3.4cm]{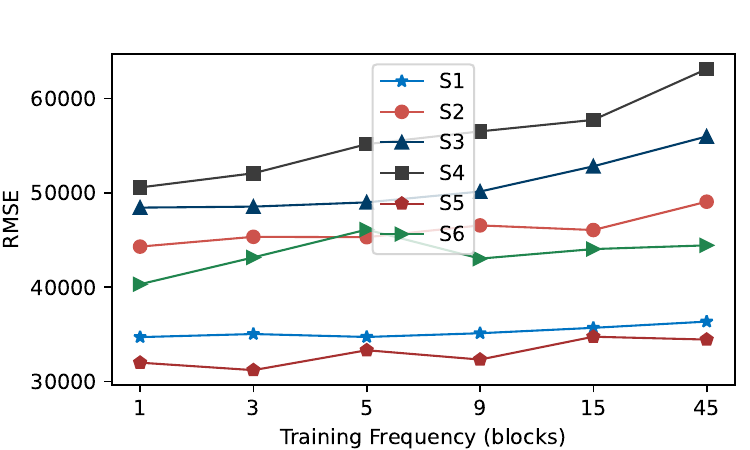}
    \caption{\protect\label{RMSE} Evaluation of training frequency on RMSE}
  \end{figure}

\section{Conclusion}
\label{secConclusion}
This work begins by documenting and analyzing previous transaction fee estimation research. Then we proposed a new neural network-based framework to analyze complex interactions from a wider range of sources, including transaction details, network features, and mempool states, in order to address the issues of inferior estimation accuracy and limited knowledge used in previous work. The effectiveness and efficiency of our suggested architecture have been demonstrated on genuine blockchain datasets.

\begin{acknowledgements}

The authors are thankful for the support from Data61, Australian Research Council Discover grants DP170104747, DP180100212, DP200103700 and National Natural Science Foundation of China grant  61872258.
\end{acknowledgements}

\bibliographystyle{spmpsci}
\bibliography{FeeEstimation}

\end{document}